# Ion-induced nanopatterning of a bacterial cellulose hydrogel

Sandra L. Arias[1]†*, Ming Kit Cheng[2]†, Ana Civantos[2], Joshua Devorkin[2], Camilo Jaramillo[2], Jean Paul Allain[1, 2] *

[1]Department of Bioengineering, University of Illinois at Urbana-Champaign, IL 61801

[2]Department of Nuclear, Plasma and Radiological Engineering, University of Illinois at Urbana-Champaign, IL 61801

† Equal contributions

*Correspondence to slarias@cornell.edu, allain@psu.edu

**Present Addresses**

S.L.A is now at the Meinig School of Biomedical Engineering, Cornell University, Ithaca, NY 14853

J. D. is now at the Pritzker School of Law, Northwestern University, Chicago, IL 60611

A. C., C. J. and J.P.A are now at The Ken and Mary Alice Lindquist Department of Nuclear Engineering, Pennsylvania State University, State College, PA 16802

**Abstract**

Hydrogels provide a solution-mimicking environment for the interaction with living systems that make them desirable for various biomedical and technological applications. Because relevant biological processes in living tissues occur at the biomolecular scale, hydrogel nanopatterning can be leveraged to attain novel material properties and functionalities. However, the fabrication of high aspect ratio (HAR) nanostructures in hydrogels capable of self-standing in aqueous environments, with fine control of the size and shape distribution, remains challenging. Here, we report the synthesis of nanostructures with a HAR in bacterial cellulose (BC) hydrogel via directed plasma nanosynthesis using argon ions. The nanostructures in BC are reproducible, stable to sterilization, and liquid immersion. Using surface characterization and semi-empirical modeling, we discovered that pattern formation was linked to the formation of graphite-like clusters composed of a mixture of C-C and C=C bonds. Moreover, our model predicts that reactive species at the onset of the argon irradiation accelerate the bond breaking of weak bonds, contributing to the formation of an amorphous carbon layer and nanopattern growth.

**Introduction**

Hydrogels are three-dimensional polymeric networks that are soft and squishy like a banana and with flexibility equivalent to that of an eye contact lens but can be stronger than a rubber band and with the capacity to retain water up to 100 times of its own weight without dissolving, making them an excellent interface with living systems[1]. Hydrogels are also ideal for biosensing applications because their hydrophilic nature reduces the non-specific interaction with proteins and cells while preserving the functionality of enzymes or other biomolecular detection probes[2]. Hydrogels have been used as multifunctional materials in a variety of biomedical and technological applications, including carriers for drug delivery systems, membranes for separation and bioanalysis[3], actuators in biosensing devices[4], magnetic composites for tissue engineering[5], and the fabrication of flexible conducting materials[6]. Hydrogels not only provide a solution-mimicking environment for the interaction with living systems, but also their swelling/deswelling volume ratio can be adjusted to respond to external stimuli such as changes in the pH, temperature, and ionic strength of the local environment[7]. Because physiologically relevant processes in living tissues occur at the biomolecular scale, hydrogel nanopatterning can be leveraged to attain novel material properties and functionalities. For example, high aspect ratio nanostructures in a hydrogel increase the surface area active for molecular detection, which leads to higher sensitivity and improved signal-to-background ratio in microfluidic and sensing devices[8,9]. Similarly, higher surface areas can deliver greater drug doses in a controlled manner compared to conventional flat surfaces[10]. Recently, it has also been shown that nanostructures can be used for the design of biofouling resistant surfaces inspired by naturally occurring bactericidal nanostructures[11].

Photolithography and replica molding techniques have been traditionally used to fabricate nanostructures in hydrogels; however, none of those approaches achieve nanostructures with dimensions below a few microns while maintaining precise control of the size and shape distribution[12]. Indeed, creating homogenous and reproducible nanosized features, with high aspect ratios, in soft materials at scale remains challenging. Nano-features fabricated on hydrogels have been shown to collapse in aqueous solutions when the aspect ratio is above six, mainly due to the capillary forces, limiting *in situ* applications[13]. Here, we report the fabrication of high aspect ratio nanostructures with graphite-like clusters in a hydrogel, i.e., bacterial cellulose (BC), that are resilient to capillary forces in aqueous environments. We provide a semi-empirical model of the ion-induced surface erosion and carbon enrichment based on the structural and chemical transformation of the fibrous hydrogel at increasing argon-ion doses using directed plasma nanosynthesis (DPNS). DPNS extracts broad-beam ions from a plasma source, which are then accelerated to the target material at low energies. This technique drives nanopatterning via several ion-driven synergistic physical and chemical mechanisms, including roughness and compositional-dependent preferential sputtering, energy-driven mass redistribution, surface diffusion, enrichment and depletion[14–16]. We demonstrate that in BC, irradiation also induces additional surface chemical changes such as bond breaking and cross-linking. Moreover, we show that a unique feature of low energy ion irradiation is that the modification occurs at the material surface and does not require an elevated temperature. As demonstrated on a considerable number of technologically relevant materials including Si[17], III-V semiconductors[18], elementary metals, and titanium alloys[19], low-energy ion beam irradiation has the potential of large-area processing, nanopattern reproducibility, and stability. We believe the nanostructures fabricated in the BC are of practical significance for the application in various fields such as polymer-based conducting materials, biosensors, optoelectronics, and anti-biofouling interfaces of industrial and clinical importance.

**Results and discussion**

Bacterial cellulose (BC) is a natural hydrogel composed of β-D-glucopyranose units linked by β(1-4) glycosidic bonds, which is produced extracellularly by *Komagataeibacter xylinum* as a biofilm (pellicle) at the air and liquid interface[20]. Pristine BC consists of interlaced ribbons with widths of 64.65 ± 12 nm (**Fig. 1A**), which are made up from smaller microfibrils, and it is estimated that there are about 46 microfibrils per ribbon and between 30 and 200 chains of cellulose in the cross-section of one microfibril[20]. The biomedical applications of the BC are diverse and include wound dressing and artificial skin substitutes, artificial cornea, urinary conduits, membranes for cardiovascular and neuroendovascular reconstruction[5,21,22], implants for dental, cartilage-meniscus and bone tissue, as well as a base material for the fabrication of nanocomposites in drug delivery systems[23].

BC was irradiated with argon ($Ar^+$) ions at low energy (1 keV), normal angle of incidence, and fluences in the range of 0.1 to $10\times10^{17}$ cm$^{-2}$. The interaction of $Ar^+$ irradiation with BC produced a marked morphological transformation that depended on the number of ions impacting the material per unit area (or fluence) (**Fig. 1A**). As described above, this transformation is different from pyrolysis of cellulose in that it is due to a number of synergistic physical and chemical mechanisms resulted from the energy and momentum transfer to surface atoms by incident ions, without the need of external energy input such as elevated temperatures, which can cause undesirable changes to the bulk. At low $Ar^+$ doses of $0.1\times10^{17}$ cm$^{-2}$, cellulose ribbons exhibited separate latent track pores with diameters of $19.27\pm2.09$ nm. At $Ar^+$ doses between 0.5 to $1\times10^{17}$ cm$^{-2}$, the latent track pores started overlapping, leaving behind opened or closed cavities that fuse with the existing porous of the material (**Fig. 1B**). Further $Ar^+$ irradiation, at fluences of $5\times10^{17}$ cm$^{-2}$, the BC showed a marked structural transformation where interlaced ribbons were not distinguishable anymore, while self-standing structures start evolving. These self-standing structures continued growing at higher fluences ($10\times10^{17}$ cm$^{-2}$), forming well-defined nanostructures with an average height of $412.99\pm8.77$ nm, and bottom and tip widths of $73.40\pm16.06$ nm and $47.97\pm10.98$ nm, respectively (**Fig. 1C**). The aspect ratio for those nanostructures is in the range of 3 to 8, with an average value of 5.6. The aspect ratio using the nanostructure's tip widths is 8.6. The nanostructures in argon-irradiated BC did not show any signs of deformation or ground collapse after autoclaving at 120C for 20 min, post-immersion in aqueous solution, and ethanol evaporation (**Fig. 1D**). Indeed, the height of the nanostructures remained unaltered after being exposed to the different treatments (no statistical difference), demonstrating its mechanical stability to resist heat and capillary forces in liquid environments (**Fig. 1D**). Only the sample exposed to ethanol evaporation exhibited a 7.5% decrease of the bottom's width compared to the control sample kept in the air ($p<0.05$). The internal morphology of the nanostructures seemed to consist of cross-bracing fibers, which could contribute to the observed mechanical stability. Similarly, the effective Young's modulus of the argon-irradiated BC obtained at the highest fluence ($10\times10^{17}$ cm$^{-2}$) increased from $4.86\pm1.05$ to $8.77\pm0.51$ MPa respect to the pristine hydrogel (**Fig. 1E**), as determined in aqueous solution (ultrapure water) using a soft material nanoindenter.

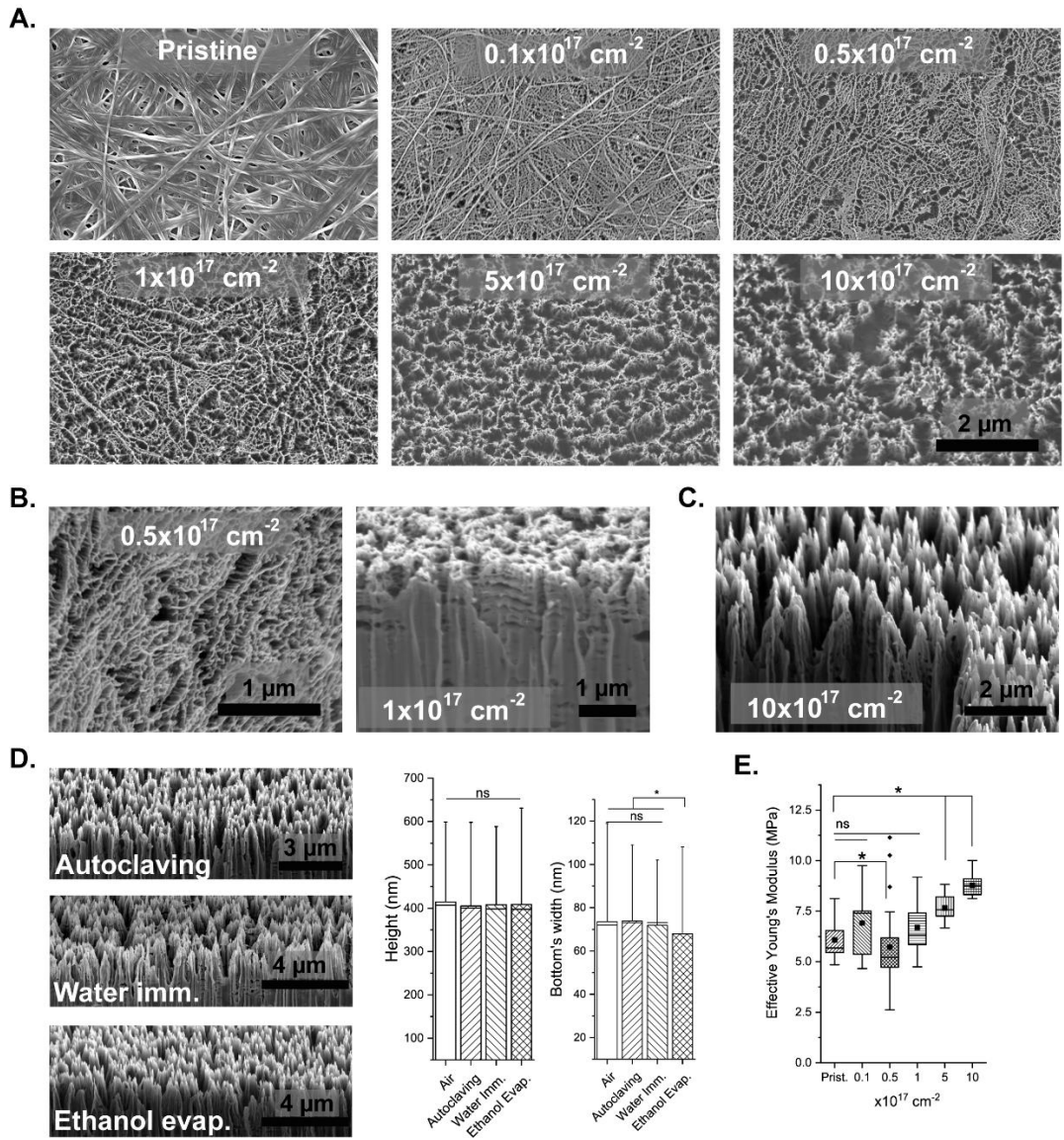

**Figure 1. Argon irradiation at high fluences induces nanostructure growth in bacterial cellulose hydrogel.** (**A**) Micrographs depicting nanopattern evolution at increasing Ar+ ion fluences (from 0.1 -10x10$^{17}$ cm$^{-2}$); (**B**) at intermedium fluence, latent track pores start overlapping forming cavities that fused with the existing porous in the material; (**C**) nanostructures in BC start forming at 5x10$^{17}$ cm$^{-2}$ and continue growing in height with further argon implantation at 10x10$^{17}$ cm$^{-2}$. (**D**) Treatment with heat by autoclaving at 120 °C for 20 min, water immersion and ethanol evaporation did not deform or ground collapse the nanostructures, as the dimensions of those are conserved after each treatment (ns: not significant, $p<0.05$). The error bar denotes the standard deviation. (**E**) The effective Young's modulus for pristine BC (prist.) and argon-irradiated

BC at fluences in the range of 0.1-10x10$^{17}$cm$^{-2}$. At the highest fluence of 10x10$^{17}$cm$^{-2}$, the effective Young's modulus increased by 44.6% compared to the pristine hydrogel (ns: not significant, *$p<$ 0.05). The middle lines in the box plots indicate the median, the box represents the interquartile range, the diamond shows the data outliners, and the square corresponds to the mean. At least 15 indentations were performed per sample in ultrapure water. Experimental data were analyzed using one-way ANOVA with Tukey post-test analysis.

XPS analysis revealed significant changes in the overall oxygen-to-carbon (O:C) ratio, as well as ratios between various functional groups involving carbon and oxygen bonds, upon irradiation. The overall O:C ratio was decreased from 0.67 to as low as 0.15 at the highest fluence of 10x10$^{17}$ cm$^{-2}$ **(Fig. 2A).** The O:C ratio drastically reduced at the beginning but became steadier after a fluence of 1x10$^{16}$ cm$^{-2}$. A closer examination of the C1s region revealed significant changes in the C-C and C-O bonding states **(Fig. 2B)**. A deconvolution of the C1s peak of the pristine BC indicated the presence of two major functional groups that exist in the BC structure, namely C-O (286.5 eV) and O-C-O (287.9 eV), plus minor contributions from C-C (284.8 eV), C=O (~288 eV, overlapped with the O-C-O peak) and COO- (289.4 eV) groups due to the presence of adventitious carbon (**Fig. S1**). After irradiation, the number of C-C groups increased significantly, and the number of C-O groups decreased (**Fig. 2C**, **Fig. S1**). The amount of C=O/O-C-O groups initially reduced at the lowest fluence, then became steady beyond that (**Fig. 2C**). The amount of COO$^-$ groups remained constant throughout the fluence regime (**Fig. 2C**). The peak also became asymmetric with a broad tail towards the higher binding energy side. Peak broadening of the C-C peak can also be seen. The overall full width at half maximum (FWHM) of the C1s peak increased from 1.44 eV to 2.06 eV initially at the lowest fluence, then decreased and became steady at around 1.6-1.7 eV at higher fluence (**Fig. S2**). A deconvolution of O1s also indicated a decrease of C-O groups, but a small increase of C=O groups (relative to the total number of oxygen functional groups) after irradiation (**Fig. 2C, Fig. S3**). A shoulder on the higher binding energy side at around 536 eV was also seen. This is probably a sodium Auger line overlapping with the O 1s peak rather than any additional oxygen-containing functional groups since there was a slight amount of sodium observed on the surface after irradiation, as indicated in supporting **Fig. S4**. It is worth noting that the keV ion-induced modifications are limited to material's topmost layer (few nanometers in thickness), as demonstrated on other organic polymers such as polystyrene (PS), poly-methyl methacrylate (PMMA) and poly(α-methylstyrene) (PαMS)[16]. Because the polymer transformation is limited to its surface, this leaves the material's bulk properties unaffected, including its capacity to absorb large amounts of water. Also, and although we irradiated dried BC pellicles, this does not mean pattern formation and the transformation of the material upon irradiation were linked to its hydration state.

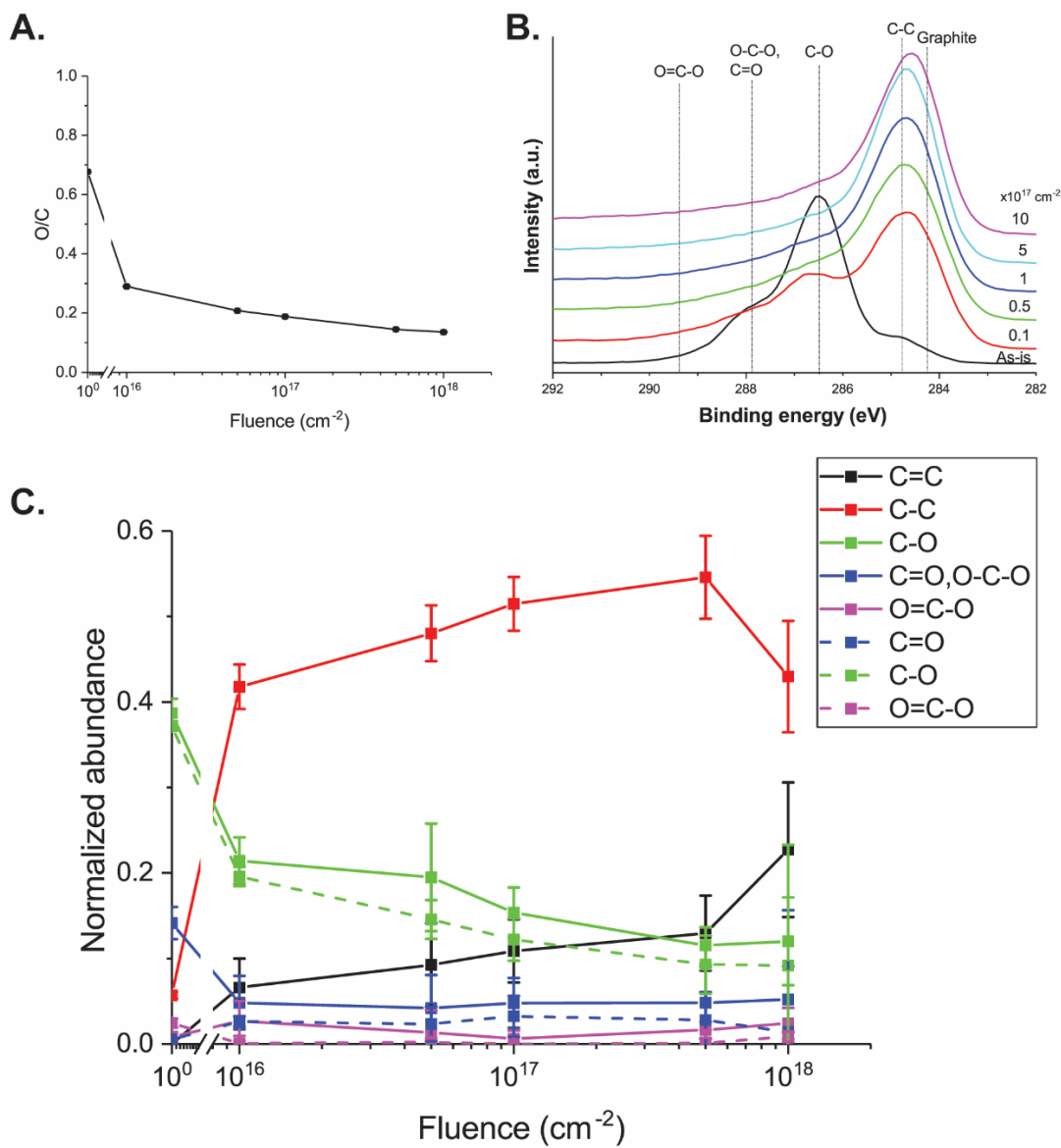

**Figure 2. Graphite-like compounds formed at the surface of BC with increasing ion implantation doses, while oxygen atoms were preferentially sputtered from the hydrogel.**

**(A)** Overall oxygen to carbon (O/C) ratio as a function of the fluence shows the decrease in the oxygen content; **(B)** X-ray photoelectron showing the changes in C1s in pristine and argon-irradiated BC at increasing ion fluences with the formation of graphite-like clusters; **(C)** Amount of different chemical bonds on the surface as a function of the Ar+ fluence. Solid lines are derived from C1s spectra and dashed lines from O1s spectra.

While there is an extensive body of literature that has examined irradiation effects on cellulose by low-temperature plasma (LTP) and electron beam (e-beam)[24–27], ion irradiation is unique in that it can induce different modifications compared to e-beam irradiation due to the higher mass of the ions compared to that of the electrons. Moreover, the changes during ion irradiation can be decoupled from those due to electrons, energetic neutrals, and radicals in an LTP, thus providing better controllability. Nevertheless, from the XPS results, some of the effects seem to be comparable. The literature mentioned above all indicated the decrease in O:C ratio and increase in the amount of C-C bonds relative to C-O, which is consistent with the trend observed in the XPS results in this work. It is understood that e-beam irradiation restructures the cellulose surface by three major processes: depolymerization, ring-opening, and preferential removal of oxygen atoms[24]. Based on the changes in O/C ratio in our work, preferential sputtering of O atoms also occurred in ion irradiation. This is also supported by TRIM simulation, which reveals a higher sputtering yield of 0.16 atoms/ion for O atoms compared to 0.04 atoms/ion for C atoms, respectively. Note that although XPS is unable to detect changes in the number of hydrogen atoms, it is expected that preferential sputtering of H atoms, i.e., dehydrogenation, a common phenomenon in ion-induced sputtering of polymers[16], also occurred. Bond-breaking of C-O bonds also occurred based on the C1s region analysis **(Fig. 2B, Fig. S1)**, implying the occurrence of ring-opening events. The amount of C=O/O-C-O bonds decreased initially at the lowest fluence and then became steady at higher fluence (**Fig. 2C**). The decrease was mainly due to the removal of O-C-O bonds because C=O bonds were not present in the pristine structure, and the O-C-O bond-breaking implies chain scission/depolymerization. The steady trend might also imply O-C-O bonds were mostly removed, and only C=O from adventitious carbon was contributing to the peak at ~288 eV. This, together with the steady trend for the COO⁻ group and the preferential removal of O atoms, also indicated the formation of C=O bonds due to irradiation were unlikely. However, the major difference observed when compared to e-beam[27] and plasma-induced modifications of cellulose[25] was the rise of peak asymmetry and broadening. Deconvolution of the broadened C-C peak from the irradiated substrates revealed the presence of both C-C (284.8 eV) and C=C (284.2-284.4) groups. The amount of C=C species gradually increased with ion fluence, with a more significant jump when the highest fluence was reached. Indeed, several other authors have indicated that heating/ion irradiation of carbon-containing compounds can lead to surface amorphization[28,29]. Amorphous carbon is composed of a mixture of C-C and C=C bonds and is graphite-like. Graphite and graphite-like compounds usually show an asymmetric C1s peak. Also, due to the disordered surface structure, peak broadening is expected to be seen. Therefore, we believe the presence of amorphous carbon on the surface causes the broadening and asymmetry, and such mechanism is not as commonly seen in e-beam or LTP exposure in which ions have low energies.

Moreover, we believe this amorphous carbon layer consisted of heavily cross-linked carbon atoms because a significant amount of O and H atoms had been removed, leaving behind carbon dangling bonds resulting from ring-opening and depolymerization. Indeed, molecular dynamics (MD) simulation of $Ar^+$ sputtering on many oxygen-containing polymers such as PMMA and poly-methyl isopropyl ketone (PMIPK) also indicated at the steady-state sputtering condition, a damaged layer of the order of a few nanometers containing mostly C atoms with a high degree of cross-linking is formed[16]. In contrast, amorphization and cross-linking were not common in e-beam or LTP exposure. The numerical simulation work by Polvi and coworkers[24] indicated that cross-linking was uncommon with up to 500 keV e-beam irradiation, with 0.05 cross-links only per recoil atom. The work by Hua and coworkers[30] on Ar plasma treatment of cellulose also did not show the formation of a cross-linked surface rich in carbon. A cross-linked layer can also explain the improved mechanical stability of the irradiated BC surface. We further confirmed the presence of amorphous carbon in the nanostructured BC by identifying the G and D bands typical of graphite via Raman spectroscopy (**supplementary text** and **Fig. S5**).

It is interesting to note that an amorphous carbon-rich surface has already formed well before nanostructure formation, leading to a possible conclusion that irradiation of BC resembles that of an amorphous carbon surface. However, there was no literature suggesting nanostructures with heights of the order of hundreds of nanometers like the ones reported in this work can be formed on pure amorphous carbon or any other elemental target at room temperature[17,31]. Existing theories are mostly built upon simple elemental targets and binary compounds such as Si[14] and III-V semiconductors[15]. Hence, they are not sufficient to explain the growth of large-scale nanostructures at room temperature, with dimensions significantly exceeding the typical depths of a keV ion-induced damaged region. Zhou and coworkers[32] found that nano-cone arrays with dimensions like those reported in this work can be grown on impurity-added Si surface. The mechanism underlying this phenomenon was attributed to the formation of metal silicide cluster regions, which acted as a local sputter shield because of its lower sputter yield. Moreover, the difference in sputter yield was suggested to lead to height variations and nanopattern growth[17,32]. However, in BC, the irradiated surface consists mostly of carbon, possibly with some residual H- and O-containing phases, which have a higher sputter yield than carbon. However, we believe this shielding mechanism can still occur via the formation of a partially cross-linked surface, as evident in our XPS analysis. In other words, our data support a scenario where the BC surface is not entirely composed of cross-linked carbon. Instead, cross-linked clusters are formed at discrete regions, which have a lower sputter yield than the non-cross-linked parts. As the ion fluence is increased, the number of cross-links and thus the area of the clusters augment, further amplifying the compositional-dependent preferential sputtering effects, and eventually leading to HAR nanostructures. Interestingly, in work performed by Zhou and coworkers[32], high temperature (>500 $^{\circ}$C) was needed to grow large nano-cone arrays on Si. Still, in the present study, we achieved similar patterns at room temperature. On the other hand, the initial roughness on a pristine BC may also contribute to the effect of roughness-dependent preferential sputtering on nanopattern formation. To further examine these proposed mechanisms and to decipher the underlying mechanisms unique to polymers in general, a separate, more in-depth work that incorporates numerical simulations such as MD or atomistic simulation will be necessary. Nevertheless, this work has shown the ability of DPNS to grow sub-micron, structurally stable structures on a large-scale BC surface while avoiding damage to the bulk.

The etching rate as a function of ion fluence was studied to gain a better understanding of the ion-induced cross-linking of the BC surface and its influence in the ion irradiation process. **Figure 3A** shows the etched thickness as a function of ion fluence. The etching process can be divided into two stages: a transient stage with an initially high sputtering rate, which decreases over time, and finally transitions to a steady-state with a constant rate. The initial sputter rate at around $3\times10^{15}$ ions/cm² fluence and the steady-state rate were 276.3 and 67.9 nm/min, respectively. The transition from the transient to steady-state occurred at ~3 x $10^{16}$ ions/cm². The average ion etching current was kept at 0.16 mA/cm². This phenomenon has also been observed in low energy (1 keV) Ar⁺ irradiation of a vast number of polymers such as polyethylene terephthalate (PET), PS, PMMA, and PαMS[33,34]. This is believed to be due to a transition to an ion-induced cross-linked surface after a specific polymer-specific fluence, which then becomes more challenging to sputter than the pristine surface. The occurrence of this transition is also supported by XPS results, which shows that a cross-linked, carbon-rich amorphous surface appears at a fluence between $1\times10^{16}$-$5\times10^{16}$ ions/cm⁻², the range in which the transition fluence estimated from the sputter rate data lies. The initially high sputter yield can also be explained by the fact that BC has a low mass density[35] ($\approx$1.25g cm⁻³) and a small surface binding energy because the polymer chains are linear and weakly bonded by van der Waal's forces. This leads to a high sputter yield initially since it is inversely proportional to atomic density times the surface binding energy. Eventually, the surface becomes carbon-rich and cross-linked, leading to a decrease in yield.

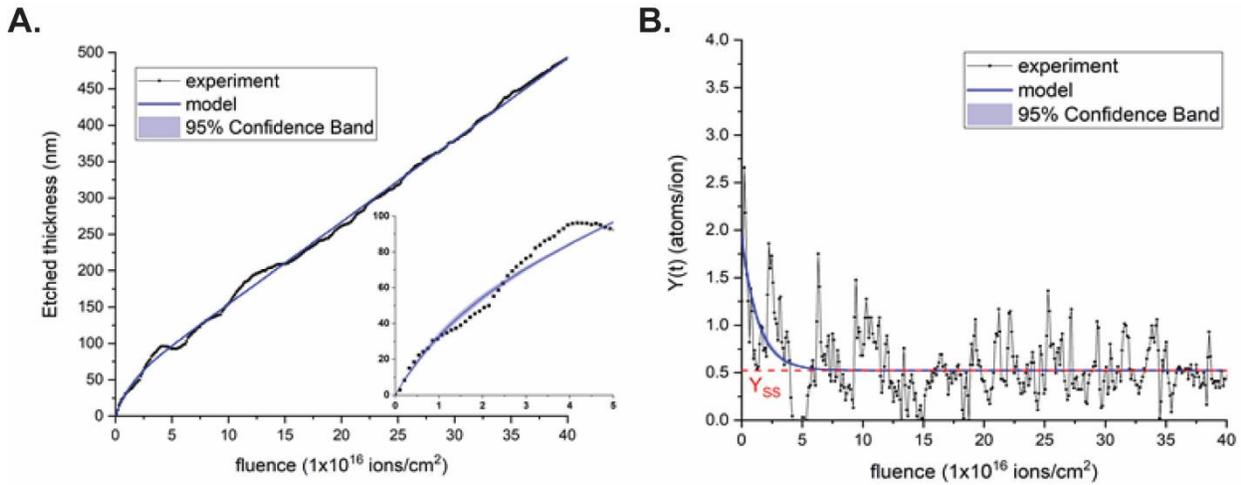

**Figure 3. Experimentally measured etched thickness as a function of ion fluence. (A)** A plot of time-dependent sputter yield as a function of ion fluence. The blue curve used to fit the data was derived from the model. Inset is a zoomed-in the region of fluence below $5\times10^{16}$ cm⁻², with the same units for the x and y-axes. **(B)** The experimental yield was derived from taking the derivative of the measured etched thickness according to equation (1). The fluctuation is likely due to the noise in QCM measurements, which is amplified after taking the derivative.

Sputtering rate is typically characterized for elemental targets by the atomic sputter yield $Y$ defined as the number of target atoms ejected from the surface per incident ion. For polymers, $Y$ is defined similarly by converting the experimentally measured sputter rate to the number of atoms removed by the relation $N\dot{h} = j_i Y$, where $\dot{h} = dh/dt$ is the sputter rate, $N$ is the atomic density of the polymer (atoms/cm$^3$), $j_i$ is incident ion flux (ions/cm$^2$·s). The steady-state sputter yield is linearly related to the Ohnishi parameter, with a constant factor empirically obtained from an experiment or from MD simulations that depend on ion energy[16,36]. In our work, to model also the non-linear part at low fluence, we introduce a time dependency to the yield $Y(t)$ as follow:

$$N\dot{h} = j_i Y(t) \quad (1)$$

$Y(t)$ is modeled with a simple function that consists of a transient and steady-state part:

$$Y(t) = Y_{ss} + Y_t(t) \quad (2)$$

$$Y_t(t) = Y_c \exp(-t/t_c) \quad (3)$$

where $Y_{ss}$ and $Y_t$ denote steady-state and transient yield, respectively. $Y_t$ has the form of an exponentially decaying function with fitting parameters $Y_c$ and $t_c$ to depict the decaying influence of the transient part and eventually being dominated by the steady-state part. A fit to the experimental data is shown as the solid line in **Fig. 3**. In the fitting procedure, $t$ is first converted to fluence $F$ by $t = F/j_i$, where $j_i$ =9.76x10$^{14}$ ions/cm$^2$.s is the average ion flux used in the experiment. It can be seen the fitting was quite good despite fluctuations between 1-15x10$^{16}$ ions/cm$^2$, with an R-square value of 0.9991. $Y_{ss}$, $Y_c$, and $t_c$ were found to be equal to 0.5235 ± 0.00087 atoms/ion, 1.4 ± 0.10 atoms/ion, and 14 ± 1.1 s, respectively.

The behavior of $Y_t$ and the physical meaning of the parameters $Y_c$ and $t_c$ can be examined more closely by plotting $Y_t$ against fluence, as shown in **Fig. 3B**. The initial yield at $t=0$ equals to $Y_c=1.924$, and the yield at $t \to \infty$ (steady-state, $Y_{ss}$) equals to 0.524. The decrease in yield is more than three folds. Comparing to several other common polymers such as PS, PMMA and PαMS, whose initial sputter yields at 1 keV $Ar^+$ are reported to be 3, 19, 7 atoms/ion respectively, and steady-state yield reported to be 1.2, 3.1 and 1.6 atoms/ion respectively[34], BC has a lower sputter yield both initially and during steady state, despite having a higher oxygen content. This may be due to preferential cross-linking rather than degradation in BC that is known to decrease the sputter yield[34], as can be seen in the early emergence of a cross-linked layer at a low fluence shown in the XPS results. A more detailed study on the sputter yield dependence on the BC polymer structure and characteristics, including irradiation of other conventional polymers within the same experimental settings, need to be carried out to decipher the difference. The physical meaning of $Y_c$ can be examined from the proposition that polymer sputtering can be divided into two sputtering mechanisms: chemical and physical sputtering[33]. Chemical sputtering involves the creation of reactive species such as free radicals and H atoms upon ion irradiation from ion-induced bond-breaking events, which then further reacts at the open bonds and create molecules that desorb from the surface. Hence chemical sputtering exhibits a much higher sputter yield than physical sputtering. In this model, since the initial sputter yield at $t = 0$ is equal to $Y_{ss} + Y_t(t = 0) = Y_{ss} + Y_c$, $Y_{ss}$ can then be regarded as the yield due to physical sputtering plus a contribution from chemical reaction of physically sputtered O atoms with C atoms governed by the Ohnishi parameter, and the constant $Y_c$ being the additional chemical sputtering yield due to the formation of free radicals. The other constant $t_c$ can be regarded as the decay constant of the influence of chemical sputtering to the overall sputter rate. The physical picture in this proposition is that at the initial stage when $t$ is small, chemical sputtering by radical reaction with the surface is the dominant mechanism with a sputter yield $Y_c$ of 2.67 times larger than physical sputtering. As irradiation continues the influence of chemical sputtering decays with a characteristic decay time constant $t_c$, due to ion-induced cross-linking and preferential removal of H and O atoms leading to an amorphous carbon-rich surface, thus reducing the surface reactivity. At the steady-state at which the surface resembles an amorphous carbon layer, only physical sputtering of C plus some residual H and O atoms at the surface contributes with a yield $Y_{ss}$. For hydrocarbons, $Y_{ss}$ is close to the physical sputter yield of amorphous carbon, and for oxygen-containing polymers, the Ohnishi parameter needs to be included to account for the presence of reactive oxygen[16]. Since sputter rate profiles are often measured as a function of fluence instead of time, $t_c$ can be multiplied by $j_i$ to obtain the decay constant for fluence, $\mathsf{F}_c$. Moreover, as many polymers exhibit similar trends in sputter rate profiles[33,34], we believe this simple semi-empirical model can also be applicable to a variety of polymers by varying the parameters $Y_{ss}$, $Y_c$, and $t_c$.

**Conclusions**

Hydrogel nanopatterning is pivotal in the fabrication of hydrogel-based electronic devices, electrochemical sensors, and platforms for the controlled release of drugs and other molecules in a variety of technological and diagnostic applications. However, hydrogel nanopatterning remains elusive because nanoscale structures with high aspect ratio are mechanically unstable in aqueous environments. Here, we fabricated nanostructures in BC with an aspect ratio in the range of 3 to 8, which can resist capillary forces in aqueous solutions by using low energy singly charged argon ions. This process is different from the pyrolysis of cellulose is that the ion-induced nanopatterning of the BC is due to several synergistic physical and chemical mechanisms resulted from the energy and momentum transfer to surface atoms by incident ions. Moreover, and in sharp contrast with pyrolysis, nanopatterning of the BC occurs only at the surface, leaving the bulk properties of the material unaffected, including its capacity to absorb large amounts of water. Surface characterization of the irradiated BC revealed significant changes in the overall oxygen-to-carbon ratio after argon irradiation, that accompanied the formation of an amorphous carbon layer composed by a mixture of C-C and C=C bonds that were graphite-like in composition. Based on the sputter rate profile, we found that the BC preferably cross-linked rather than degraded. We identified two stages during the crosslinking of the BC: a transient and a steady-state stage. An exponential decay function describes the transient stage with a high sputter yield dominated by radical species (chemical sputtering), which is associated with the early emergence of the amorphous carbon. On the contrary, the steady-state stage was dominated by physical sputtering with a low sputter rate and promotes nanopattern growth by creating new radical species below the already cross-linked material. The nanostructures fabricated in the BC are of practical significance for the application in various fields such as polymer-based conducting materials, biosensors, optoelectronics, and anti-biofouling interfaces of industrial and clinical importance. Our results also indicate that a graphite-like surface can be obtained in cellulose without increasing the temperature of the material, which is of practical interest in industrial settings.

**Methods**

**Preparation and irradiation of the bacterial cellulose (BC).** Never-dried BC pellicles (Jenpolymers, Germany), were cut in pieces of 0.25 $cm^2$ and dried at room temperature on glass slides of the same size. Air-dried BC samples were then inserted in a vacuum chamber and irradiated with argon ions ($Ar^+$) at fluences of 0.1, 0.5, 1, 5, and $10 \times 10^{17}$ $cm^{-2}$, respectively. An ion energy of 1 keV at normal incidence was used for all the experiments. The base pressure of the vacuum chamber was $5 \times 10^{-6}$ Pa, and ion beam irradiation was performed at a gas pressure of $2 \times 10^{-4}$ Pa. The operating current was kept between 0.2-0.3 mA.

**Scanning electron microscopy and focused ion beam.** Pristine BC and argon-irradiated BC were morphologically characterized by scanning electron microscopy (SEM) (4800 Hitachi) and focused ion beam (FIB) (Thermo Scios 2 Dual-Beam SEM/FIB). For both SEM and FIB, substrates were sputtered with an Au-Pd thin film at a deposition current of current of 20 mA during 40 s, yielding a total film thickness of 100 Å. SEM images were acquired using a voltage of 5 keV and a probe current of 5 mA. For rough cross-sections using the FIB, a rectangular cross section of 25x15 µm in size and 3 µm in depth were made using an ion beam current of 3 nA. Cleaning cross sections were made using an ion beam current of 0.1 nA. Images were taken using a voltage of 2 kV and probe current of 0.1 nA, dwell time of 10µs, and line integration of 10. To determine the mechanical stability of the nanostructures in the Ar-irradiated BC, experimental samples were submerged either in water and ethanol overnight or autoclaved at 120 °C for 20 min. Subsequently, samples were allowed to dry at room temperature and examined by SEM/FIB. ImageJ was used for spatial calibration to measure the pore diameters, nanopillar height and diameters. Nanopillar height and widths were measured with at least 4 different SEM images per condition, and 50 nanostructures were measured per image. The aspect ratios were then calculated by dividing the height by the diameter at the bottom of the nanostructures.

**Young's modulus determination**. The Optics11 Piuma soft material indenter was used to determine Young's modulus of pristine and argon-irradiated BC for all the fluence regime. All the materials were indented in aqueous solution (ultrapure water) with a spherical glass bead of 8 μm in diameter and stiffness of 46.900 N/m. Between 15 to 25 indentations were performed per sample, with a separation of 50 μm between indentation points.

**X-ray photoelectron spectroscopy.** Surface chemistry changes were examined using X-ray photoelectron spectroscopy (XPS) (Kratos Axis ULTRA system). A monochromated Al Kα (1486.6 eV) x-ray source was used. The base pressure during the analysis was better than $3 \times 10^{-8}$ Torr. The instrument was also equipped with a low energy electron flood gun for charge compensation. A survey scan plus region scans corresponding to C1s and O1s photoelectron peaks were performed. A total of 2 survey scans and 10 region scans per region were acquired. A pass energy of 160 eV and 40 eV was used in the survey region scans, respectively. The binding energy scale was referenced to the C1s peak corresponding to C-C bond at 284.8 eV. Peak deconvolution and calculation of atomic fractions were performed using CasaXPS. A Shirley synthetic background was used. For each chemical species in its respective region, a convoluted Gaussian-Lorentzian function GL(30) was used for peak fitting. For the C=C species, an asymmetric tail was added to the Gaussian-Lorentzian function to model the asymmetry. Error analysis of the peak-fitting and atomic ratios were performed using a built-in Monte Carlo procedure in CasaXPS.

**Raman confocal microscopy**. Experimental samples were chemically characterized by Raman confocal microscopy using Nanophoton-11. Raman scattering was done by exciting with a 532 nm near-infrared diode laser with 10x and 50x objective lenses. Data acquisition covered the spectral range from 250 nm to 3600 nm in two steps, with 300 nm of lateral resolution and 500 nm of vertical resolution.

**Sputter rate measurements using a quartz crystal microbalance (QCM).** Four pellicles of never dried BC (5 mm in diameter) were mixed with 1 mL of ultrapure water and pulped using a tissue-disintegrator operating at 5000 rpm for 5 min. The pulped BC was then deposited on a clean 6 MHz AT cut gold coated crystal (Telemark, WA) and dried overnight at room temperature. The BC-coated crystal was then mounted on a QCM sensing head connected to an INFICON SQC-310 Thin Film Deposition Controller, which was used to determine the amount of removed material as a function of the change in frequency. The amount of removed material as a function of the ion fluence was collected using an SQC-310 Comm Software V6.43.

## References

(1) Ullah, F.; Othman, M. B. H.; Javed, F.; Ahmad, Z.; Akil, H. M. Classification, Processing and Application of Hydrogels: A Review. *Mater. Sci. Eng. C* **2015**, 414–433.

(2) Buenger, D.; Topuz, F.; Groll, J. Hydrogels in Sensing Applications. *Prog. Polym. Sci.* **2012**, *37,* 1678–1719.

(3) Nagase, K.; Okano, T. Thermoresponsive-Polymer-Based Materials for Temperature-Modulated Bioanalysis and Bioseparations. *J. Mat. Chem. B* **2016**, *4*, 6381–6397.

(4) Lee, B. P.; Lin, M. H.; Narkar, A.; Konst, S.; Wilharm, R. Modulating the Movement of Hydrogel Actuator Based on Catechol-Iron Ion Coordination Chemistry. *Sensor Actuat. B: Chem.* **2015**, *206*, 456–462.

(5) Arias, S. L.; Shetty, A.; Devorkin, J.; Allain, J. P. Magnetic Targeting of Smooth Muscle Cells *In vitro* Using a Magnetic Bacterial Cellulose to Improve Cell Retention in Tissue-Engineering Vascular Grafts. *Acta Biomater.* **2018**, *77*, 172–181.

(6) Pan, L. *et al*. Hierarchical Nanostructured Conducting Polymer Hydrogel with High Electrochemical Activity. *Proc. Natl. Acad. Sci. U.S.A.* **2012**, *109*, 9287–9292.

(7) Peppas, N. A.; van Blarcom, D. S. Hydrogel-Based Biosensors and Sensing Devices for Drug Delivery. *J. Control. Release* **2016**, *240*, 142–150.

(8) Chen, G.; McCandless, G. T.; McCarley, R. L.; Soper, S. A. Integration of Large-Area Polymer Nanopillar Arrays into Microfluidic Devices Using in Situ Polymerization Cast Molding. *Lab Chip* **2007**, *7*, 1424–1427.

(9) Mesch, M.; Zhang, C.; Braun, P. v.; Giessen, H. Functionalized Hydrogel on Plasmonic Nanoantennas for Noninvasive Glucose Sensing. *ACS Photonics* **2015**, *2*, 475–480.

(10) Anandan, V.; Rao, Y. L.; Zhang, G. Nanopillar Array Structures for Enhancing Biosensing Performance. *Int. J. Nanomedicine* **2006**, *1*, 73–79.

(11) Lin, N.; Berton, P.; Moraes, C.; Rogers, R. D.; Tufenkji, N. Nanodarts, Nanoblades, and Nanospikes: Mechano-Bactericidal Nanostructures and Where to Find Them. *Adv. Colloid Interface Sci.* **2018**, *252*, 55–68.

(12) Ma, S.; Yu, B.; Pei, X.; Zhou, F. Structural Hydrogels. *Polymer* **2016**, *98*, 516–535.

(13) Zhang, Y.; Lo, C. W.; Taylor, J. A.; Yang, S. Replica Molding of High-Aspect-Ratio Polymeric Nanopillar Arrays with High Fidelity. *Langmuir* **2006**, *22*, 8595-8601.

(14) Keller, A.; Facsko, S. Ion-Induced Nanoscale Ripple Patterns on Si Surfaces: Theory and Experiment. *Materials* **2010**, *3*, 4811–4841.

(15) El-Atwani, O.; Norris, S. A.; Ludwig, K.; Gonderman, S.; Allain, J. P. Ion Beam Nanopatterning of III-V Semiconductors: Consistency of Experimental and Simulation Trends within a Chemistry-Driven Theory. *Sci. Rep.* **2015**, *5*, 1–13.

(16) Choudhary, G. K.; Végh, J. J.; Graves, D. B. Molecular Dynamics Simulations of Oxygen-Containing Polymer Sputtering and the Ohnishi Parameter. *J. Phys. D Appl. Phys.* **2009**, *42*, 242001.

(17) Muñoz-García, J.; Vázquez, L.; Castro, M.; Gago, R.; Redondo-Cubero, A.; Moreno-Barrado, A.; Cuerno, R. Self-Organized Nanopatterning of Silicon Surfaces by Ion Beam Sputtering. *Mater. Sci. Eng. R Rep.* **2014**, 1–44.

(18) Facsko, S.; Dekorsy, T.; Koerdt, C.; Trappe, C.; Kurz, H.; Vogt, A.; Hartnagel, H. L. Formation of Ordered Nanoscale Semiconductor Dots by Ion Sputtering. *Science* **1999**, *285*, 1551–1553.

(19) Civantos, A.; Barnwell, A.; Shetty, A. R.; Pavón, J. J.; El-Atwani, O.; Arias, S. L.; Lang, E.; Reece, L. M.; Chen, M.; Allain, J. P. Designing Nanostructured Ti6Al4V Bioactive Interfaces with Directed Irradiation Synthesis toward Cell Stimulation to Promote Host-Tissue-Implant Integration. *ACS Biomater. Sci. Eng.* **2019**, *5*, 3325–3339.

(20) Klemm, D.; Heublein, B.; Fink, H. P.; Bohn, A. Cellulose: Fascinating Biopolymer and Sustainable Raw Material. *Angew. Chem.* **2005**, *44,* 3358–3393.

(21) Arias, S. L.; Shetty, A. R.; Senpan, A.; Echeverry-Rendón, M.; Reece, L. M.; Allain, J. P. Fabrication of a Functionalized Magnetic Bacterial Nanocellulose with Iron Oxide Nanoparticles. *J. Vis. Exp.* **2016**, *111,* e52951.

(22) Echeverry-Rendon, M.; Reece, L. M.; Pastrana, F.; Arias, S. L.; Shetty, A. R.; Pavón, J. J.; Allain, J. P. Bacterial Nanocellulose Magnetically Functionalized for Neuro-Endovascular Treatment. *Macromol. Biosci.* **2017**, *17*, 1600382.

(23) Rajwade, J. M.; Paknikar, K. M.; Kumbhar, J. v. Applications of Bacterial Cellulose and Its Composites in Biomedicine. *Appl. Microbiol. Biotechnol.* **2015**, *99,* 2491–2511.

(24) Polvi, J.; Nordlund, K. Low-Energy Irradiation Effects in Cellulose. *J. Appl. Phys.* **2014**, *115*

(25) Carlsson, C. M. Gilbert.; Stroem, Goeran. Reduction and Oxidation of Cellulose Surfaces by Means of Cold Plasma. *Langmuir* **1991**, *7*, 2492–2497.

(26) Pertile, R. A. N.; Andrade, F. K.; Alves, C. B.; Gama, M. Surface Modification of Bacterial Cellulose by Nitrogen-Containing Plasma for Improved Interaction with Cells. *Carbohyd. Polym.* **2010**, *82*, 692–698.

(27) Eo, M. Y.; Fan, H.; Cho, Y. J.; Kim, S. M.; Lee, S. K. Cellulose Membrane as a Biomaterial: From Hydrolysis to Depolymerization with Electron Beam. *Biomater. Res.* **2016**, *20,* 16.

(28) Takahiro, K. *et al*. Amorphization of Carbon Materials Studied by X-Ray Photoelectron Spectroscopy. *Nucl. Instrum. Methods Phys. Res. B* **2006**, *242*, 445-447.

(29) Ramm, M.; Ata, M.; Brzezinka, K.; Gross, T.; Unger, W. Studies of Amorphous Carbon Using X-Ray Photoelectron Spectroscopy, near-Edge X-Ray-Absorption Fine Structure and Raman Spectroscopy. *Thin Solid Films* **1999**, *354*, 106-110

(30) Hua, Z. Q.; Sitaru, R.; Denes, F.; Young, R. A. Mechanisms of Oxygen- and Argon-RF-Plasma-Induced Surface Chemistry of Cellulose. *Plasma. Polym.* **1997**, *2* (3), 199–224.

(31) Bobes, O.; Hofsäss, H.; Zhang, K. Neon Ion Beam Induced Pattern Formation on Amorphous Carbon Surfaces. *AIP Adv.* **2018**, *8*, 025205.

(32) Zhou, J.; Hildebrandt, M.; Lu, M. Self-Organized Antireflecting Nano-Cone Arrays on Si (100) Induced by Ion Bombardment. *J. Appl. Phys.* **2011**, *109*, 053513.

(33) Große-Kreul, S.; Corbella, C.; von Keudell, A. Chemical and Physical Sputtering of Polyethylene Terephthalate (PET). *Plasma Processes Polym.* **2013**, *10* (3), 225–234.

(34) Zekonyte, J.; Zaporojtchenko, V.; Faupel, F. Investigation of the Drastic Change in the Sputter Rate of Polymers at Low Ion Fluence. *Nucl. Instrum. Methods Phys. Res. B* **2005**, *236*, 241-248

(35) Lee, K.-Y.; Blaker, J. J.; Bismarck, A.; Campus, S. K. Improving the Properties of Nanocellulose/Polylactide Composites by Esterification of Nanocellulose. Can It Be Done? In: *17th international conference on composite materials. December 1, Edinburgh, UK* **2009**. https://pdfs.semanticscholar.org/d93a/ba3666997fce1a5a43c099c85e88e93106d1.pdf (accessed Jan. 1, 2020)

(36) Gokan, H.; Esho, S.; Ohnishi, Y. Dry Etch Resistance of Organic Materials. *J. Electrochem. Soc.* **1983**, *130*, 143–146.

**Acknowledgments:** Material characterization was carried out in the Materials Research Laboratory Central Research Facilities at the University of Illinois at Urbana-Champaign. Authors thank Z. Kyon for his invaluable help and technical support on performing part of the irradiations.

**Competing interests:** Authors declare no competing interests.

**Data and materials availability:** All data is available in the main text or the supplementary materials.

**Supplementary Materials:**

Supplementary Text

Figures S1-S5

References (37-40)

**Table of Contents/Abstract Graphics**

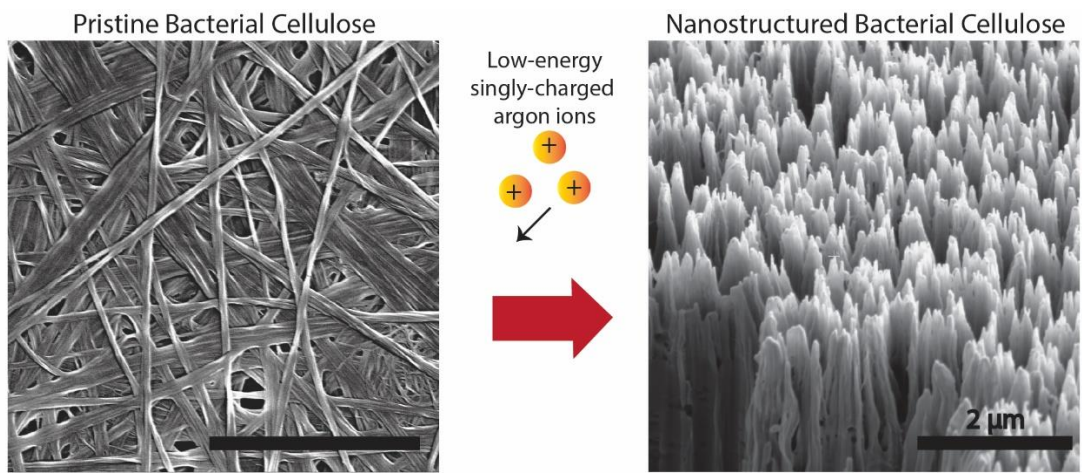

For Table of Contents Only